\documentclass[aps,prl,reprint,superscriptaddress]{revtex4-1}

\usepackage{graphicx}
\usepackage{dcolumn}
\usepackage{amsmath}

\begin{document}

\title{Evidence of a nonequilibrium distribution of quasiparticles in the microwave response of a superconducting aluminium resonator} 

\author{P.J. de Visser}
\email{p.j.devisser@tudelft.nl}

\affiliation{SRON Netherlands Institute for Space Research, Sorbonnelaan 2, 3584 CA Utrecht, The Netherlands}
\affiliation{Kavli Institute of NanoScience, Faculty of Applied Sciences, Delft University of Technology, Lorentzweg 1, 2628 CJ Delft, The Netherlands}

\author{D.J. Goldie}
\affiliation{Cavendish Laboratory, Cambridge University, JJ Thomson Avenue, Cambridge CB3 0HE, United Kingdom}

\author{P. Diener}

\affiliation{SRON Netherlands Institute for Space Research, Sorbonnelaan 2, 3584 CA Utrecht, The Netherlands}

\author{S. Withington}
\affiliation{Cavendish Laboratory, Cambridge University, JJ Thomson Avenue, Cambridge CB3 0HE, United Kingdom}

\author{J.J.A. Baselmans}

\affiliation{SRON Netherlands Institute for Space Research, Sorbonnelaan 2, 3584 CA Utrecht, The Netherlands}

\author{T.M. Klapwijk}
\affiliation{Kavli Institute of NanoScience, Faculty of Applied Sciences, Delft University of Technology, Lorentzweg 1, 2628 CJ Delft, The Netherlands}

\date{\today}

\begin{abstract}

In a superconductor absorption of photons with an energy below the superconducting gap leads to redistribution of quasiparticles over energy and thus induces a strong non-equilibrium quasiparticle energy distribution. We have measured the electrodynamic response, quality factor and resonant frequency, of a superconducting aluminium microwave resonator as a function of microwave power and temperature. Below 200 mK, both the quality factor and resonant frequency decrease with increasing microwave power, consistent with the creation of excess quasiparticles due to microwave absorption. Counterintuitively, above 200 mK, the quality factor and resonant frequency increase with increasing power. We demonstrate that the effect can only be understood by a non-thermal quasiparticle distribution.

\end{abstract}

\maketitle

A superconductor can be characterised by the density of states, which exhibits an energy gap due to Cooper pair formation, and the distribution function of the electrons, which in thermal equilibrium is the Fermi-Dirac distribution. When a superconductor is driven by an electromagnetic field, nonlinear effects in the electrodynamic response can occur, which are usually assumed to be due to a change in the density of states, the so called pair breaking mechanism \footnote{This pair breaking mechanism is different from direct Cooper pair breaking by e.g. photons.}. These nonlinear effects can be described along the lines of a current dependent superfluid density $n_s(T,j)\propto n_s(T)\left[1-(j/j_c)^2\right]$, where $j$ is the actual current density, $j_c$ the critical current density and $T$ the temperature. Observations such as the nonlinear Meissner effect \cite{syip1992} and nonlinear microwave conductivity \cite{cchin1992,beom2012} can be explained by a broadening of the density of states and a decreased $n_s$. The quasiparticles are assumed to be in thermal equilibrium and a Fermi-Dirac distribution $f(E)=1/(\exp(E/k_BT)+1)$ is assumed, with $E$ the quasiparticle energy and $k_B$ Boltzmann's constant. 

\begin{figure}

\includegraphics[width=0.99\columnwidth]{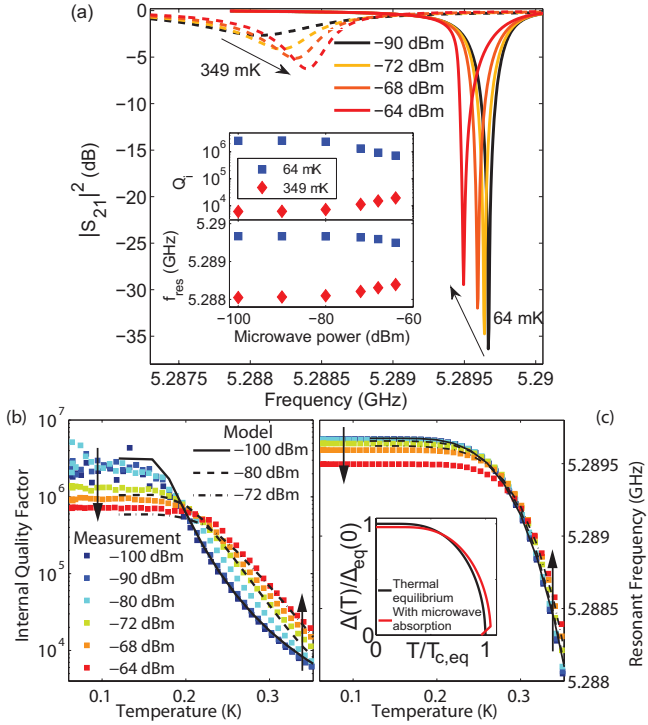}
\caption{\label{fig:S21Qf}(color online). (a) The measured microwave transmission $|S_{21}|^2$ as a function of frequency for sample A. The solid lines are taken for four different microwave readout powers ($P_{read}$) at a temperature of 64 mK. The dashed lines are taken at 349 mK. The same colour coding applies. The arrows indicate increasing $P_{read}$. The inset shows the internal quality factor and resonant frequency as determined from the $S_{21}$ measurements as a function of $P_{read}$. (b),(c) The measured internal quality factor and resonant frequency as a function of temperature for various $P_{read}$ (the same legend applies). The arrows indicate increasing $P_{read}$. Simulation results are shown as lines. The inset is a representation of the effect of microwave absorption on the energy gap $\Delta$. The temperature is normalised to the equilibrium $T_c$.}

\end{figure}

Here we demonstrate that a microwave field also has a strong effect on $f(E)$ in the superconductor, and induces a nonlinear response. We present measurements of the electrodynamic response, quality factor and resonant frequency, of an Al superconducting resonator (at 5.3 GHz) as a function of temperature and microwave power at low temperatures $T_c/18<T<T_c/3$. The response measurements, complemented with quasiparticle recombination time measurements, are explained consistently by a model based on a microwave-induced non-equilibrium $f(E)$. \\ Redistribution of quasiparticles \cite{jchang1977,jwolter1981} due to microwave absorption \cite{geliashberg1970,*bivlev1973} has been shown earlier to cause enhancement of the critical current \cite{tklapwijk1976,*tklapwijk1977}, the critical temperature ($T_c$) and the energy gap \cite{rhorstman1981}. These enhancement effects are most pronounced close to $T_c$ and were observed for temperatures $T>0.8T_c$. A representation of gap suppression and gap enhancement is shown in the inset to Fig. \ref{fig:S21Qf}b \cite{tklapwijk1977}. The consequences of the redistribution of quasiparticles for the electrodynamic response were only studied theoretically for $T>0.5T_c$ \cite{ssridhar1980}. Redistribution of quasiparticles also explains \cite{dgoldie2013} the microwave power dependent number of quasiparticles in microwave resonators at low temperatures, which we have recently measured \cite{pdevisser2012b}. These quasiparticles impose a limit for detectors for astrophysics based on microwave resonators \cite{pday2003,jzmuidzinas2012}. Related phenomena have been reported in superconducting-normal metal devices \cite{fchiodi2009,*pvirtanen2010}, terahertz pulse experiments \cite{mbeck2013} and holographic superconductivity \cite{nbao2011}. 

To measure the microwave response, microwave resonators were patterned into a 60 nm thick Al film, which was sputter deposited on a sapphire substrate. $T_c$ was measured to be 1.17 K, from which the energy gap at zero temperature is taken to be $\Delta = 1.76k_BT_c=177$ $\mu$eV. The low temperature resistivity was 0.9 $\mu\Omega$cm. The film was patterned by wet etching into distributed, half-wavelength, coplanar waveguide resonators, which are capacitively coupled to a transmission line. With readout power, $P_{read}$, we will mean the incident microwave power on the through transmission line. The presented measurements were performed on a resonator with a length of 9.84 mm and a central strip volume of 1770 $\mu$m$^3$ (sample A). Sample B is similar and will be introduced later. Further details are provided in the supplementary material \cite{supinf}. The half-wavelength geometry was chosen because it has an isolated central strip, which prevents quasiparticle outdiffusion. The samples were cooled in a pulse tube precooled adiabatic demagnetization refrigerator. Care was taken to make the sample stage light-tight as described in Ref. \onlinecite{jbaselmans2012}, which is crucial to eliminate excess quasiparticles due to straylight. The complex transmission $S_{21}$ of the microwave circuit was measured with a vector network analyser. The microwave signal was amplified at 4 K with a high electron mobility transistor amplifier and with a room temperature amplifier.

We have measured the microwave transmission $S_{21}$ for various $P_{read}$ as a function of temperature. A selection of resonance curves is shown in Fig. \ref{fig:S21Qf}a. We kept $P_{read}$ below the bifurcation regime \cite{pdevisser2010,lswenson2013}. By fitting a Lorentzian curve to the resonance curve, we extracted the resonant frequency ($f_{res}$) and the internal quality factor ($Q_i$) \cite{supinf}, which are plotted for 64 and 349 mK as a function of $P_{read}$ in the inset in Fig. \ref{fig:S21Qf}a \footnote{At 64 mK, a $P_{read}$ of -64 (-100) dBm leads to a stored energy of 0.55 fJ (0.11 aJ), corresponding to 1.6$\times 10^8$ (3.1$\times 10^4$) photons.}. $Q_i$ is higher when the resonance curve is deeper. $Q_i$ and $f_{res}$ are shown for several microwave powers as a function of temperature in Fig. \ref{fig:S21Qf}b and c. Two distinct regimes appear. At low temperatures both $Q_i$ and $f_{res}$ decrease with increasing microwave power, which is consistent with a higher effective electron temperature. At the highest temperatures however, both $Q_i$ and $f_{res}$ increase with increasing power, which contradicts with a heating model \cite{pdevisser2010} and also cannot be explained by a pair-breaking effect where the density of states broadens due to the current \cite{snam1967}. The pair-breaking mechanism would induce a downward frequency shift without dissipation \cite{beom2012} and might play a role at the highest $P_{read}$ at the lowest temperatures.

\begin{figure}

\includegraphics[width=0.99\columnwidth]{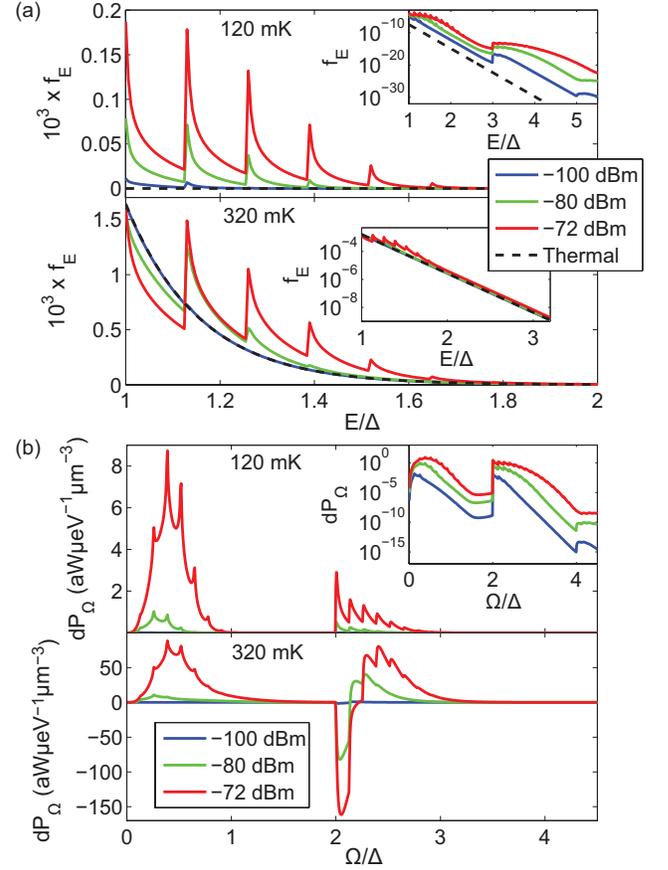}
\caption{\label{fig:distributions}(color online). (a) The calculated quasiparticle distribution as a function of normalised energy. The two different panels are for temperatures of 120 and 320 mK. The insets show the same distributions on a log-scale. (b) The phonon power flow from the film to the substrate as a function of normalised energy, for 120 mK (the inset shows the same lines on a log-scale) and 320 mK. $dP$ is zero in thermal equilibrium. }

\end{figure}

We have modelled the effect of absorption of microwave photons on the quasiparticle distribution function, $f(E)$, by using a set of kinetic equations. Absorption of a microwave photon with energy $\hbar\omega$ causes quasiparticles at an energy $E$ to move to an energy $E+\hbar\omega$. The rate with which quasiparticles at energy $E$ absorb photons with energy $\hbar\omega$ can be described with an injection term $I_{qp}(E,\omega)$ \cite{geliashberg1970,*bivlev1973}, which is given by
\begin{eqnarray}
	I_{qp}(E,\omega) &=& 2B\left[h_1(E,E+\hbar\omega)\left(f(E+\hbar\omega)-f(E)\right)\right. \nonumber\\*
&-&\left. h_1(E,E-\hbar\omega)\left(f(E)-f(E-\hbar\omega)\right)\right],
\label{eq:injection}
\end{eqnarray}
with $h_1(E,E') = \left(1+\frac{\Delta^2}{EE'}\right)\rho(E')$. $\rho(E)$ is the density of states, which is given by $\rho(E) = E/\sqrt{E^2-\Delta^2}$. $B$ relates the injection rate to the microwave field strength \cite{dgoldie2013,jmooij1983}. The thus created change in $f(E)$, is counteracted by electron-phonon scattering and quasiparticle recombination, which depend both on $f(E)$ and on $n(\Omega)$, the phonon distribution in the film ($\Omega$ is the phonon energy). In steady state the microwave power that is absorbed by the quasiparticle system is transported through the phonon system of the film and is released in the phonon system of the substrate, the heat bath. We solve the full nonlinear kinetic equations as presented in Ref. \onlinecite{jchang1977}, together with Eq. \ref{eq:injection}, in steady state, $df(E)/dt = dn(\Omega)/dt = 0$ for all energies, with a self-consistency equation for $\Delta$, given by
\begin{equation}
\frac{1}{N_0V_{BCS}}=\int_{\Delta}^{\Omega_D}\frac{1-2f(E)}{\sqrt{E^2-\Delta^2}}dE,
\label{eq:selfconsistent}
\end{equation}
with $N_0$ the single spin density of states at the Fermi level, $\Omega_D$ the Debye energy and $V_{BCS}$ the effective pairing potential. The numerical procedure is explained in Ref. \onlinecite{dgoldie2013}. 

The complex conductivity $\sigma = \sigma_1 - i\sigma_2$, describing the response of both Cooper pairs and quasiparticles to a time-varying electric field with $\hbar\omega<2\Delta$, is given by \cite{dmattis1958}
\begin{eqnarray}
\frac{\sigma_1}{\sigma_N} (\omega) &=& \frac{2}{\hbar\omega}\int^\infty_{\Delta}[f(E)-f(E+\hbar\omega)]g_1(E)dE , \label{eq:sigmaone} \\
\frac{\sigma_2}{\sigma_N}(\omega) &=& \frac{1}{\hbar \omega}\int^{\Delta}_{\Delta-\hbar\omega}[1-2f(E+\hbar\omega)]g_2(E)dE ,
\label{eq:sigmatwo}
\end{eqnarray}
where $g_1(E) = h_1(E,E+\hbar\omega)\rho(E)$ and $g_2(E) = h_1(E,E+\hbar\omega)E/\sqrt{\Delta^2-E^2}$. $\sigma_N$ is the normal-state conductivity and $\omega$ the angular frequency. Eqs. \ref{eq:sigmaone} and \ref{eq:sigmatwo} show the role of $f(E)$ in determining the conductivity. In a microwave resonator $f_{res}$ is proportional to the imaginary part of the conductivity, $\sigma_2$, and $Q_i$ is proportional to $\sigma_2/\sigma_1$, which connects these observables to $f(E)$.

Since $I_{qp}$ is proportional to the field strength squared, we need to know the microwave field in the resonator for a certain $P_{read}$. We solve this problem by using the absorbed microwave power in the quasiparticle system, $P_{abs}$. For the experiment $P_{abs}$ can be calculated by 
\begin{equation}
P_{abs} = \frac{P_{read}}{2}\frac{4Q^2}{Q_iQ_c}\frac{Q_i}{Q_{i,qp}}.
\label{eq:Pabs}
\end{equation}
The loaded quality factor $Q$ is given by $Q = \frac{Q_iQ_c}{Q_c+Q_i}$ and $Q_c$ is the coupling quality factor. $Q_c=\pi/(\omega C_g Z_0)^2$, with $C_g$ the coupling capacitance and $Z_0$ the characteristic impedance of the transmission line. See the supplementary material \cite{supinf} for a derivation. Since $Q_i$ depends strongly on temperature, $P_{abs}$ is more than an order of magnitude higher at 300 mK (where $Q_i=Q_c$) than at 100 mK \cite{supinf}, which is a crucial ingredient to model the measurements in Fig. \ref{fig:S21Qf}. The factor $Q_i/Q_{i,qp}$ in Eq. \ref{eq:Pabs} arises when $Q_i$ is not limited by quasiparticle dissipation. Here we take $Q_i/Q_{i,qp}=1$. $P_{abs}$ is calculated per unit volume, where the volume is taken to be twice that of the central strip of the resonator, to roughly account for the groundplane of the waveguide, in which power will be absorbed as well. In the calculations we adjust the constant $B$ in Eq. \ref{eq:injection}, such that $P_{abs} = 4N_0\int_{\Delta}^{\infty}I_{qp}E\rho(E)dE$. 

The simulations where performed for a frequency of 5.57 GHz. The resulting non-equilibrium quasiparticle distributions are shown in Fig. \ref{fig:distributions}a for three readout powers for temperatures of 120 and 320 mK. A structure with sharp peaks at multiples of $\hbar\omega/\Delta$ shows up due to microwave photon absorption. At 120 mK, the driven distribution exceeds the thermal distribution at the bath temperature for all energies, meaning that excess quasiparticles are created. At 320 mK, the number of quasiparticles only increases a little at higher power, but quasiparticles are taken away from energies $\Delta<E<\Delta+\hbar\omega$.

In Fig. \ref{fig:distributions}b we show the corresponding phonon power flow to the heat bath: $dP(\Omega) = 3N_{ion}D(\Omega)\Omega[n(\Omega)-n_{sub}(\Omega,T_{bath})]/\tau_{esc}$. The phonons in the film have a non-equilibrium distribution $n(\Omega)$. Phonons can escape to the substrate, the bath. The phonon distribution in the substrate $n_{sub}(\Omega)$ is assumed to have a Bose-Einstein distribution at the bath temperature $T_{bath}$. $\tau_{esc} = 0.17$ ns is the phonon escape time, calculated for Al on sapphire using the acoustic mismatch model \cite{skaplan1979}. $N_{ion}$ is the number of ions per unit volume and $D(\Omega)=3\Omega^2/\Omega_D^3$ is the phonon density of states. Fig. \ref{fig:distributions}b shows strong non-equilibrium behaviour as well, with peaks at multiples of $\hbar\omega$. Phonons at $\Omega<2\Delta$ arise due to scattering. At energies $\Omega>2\Delta$ phonons due to both recombination and scattering occur. At 320 mK, we observe phonon transport out of the film, but also into the film ($dP(\Omega)<0$ at energies $\Omega>2\Delta$). This is a consequence of the depletion of $f(E)$ for energies $\Delta<E<\Delta+\hbar\omega$ (Fig. \ref{fig:distributions}a) \cite{agulian1982}.

\begin{figure}

\includegraphics[width=0.99\columnwidth]{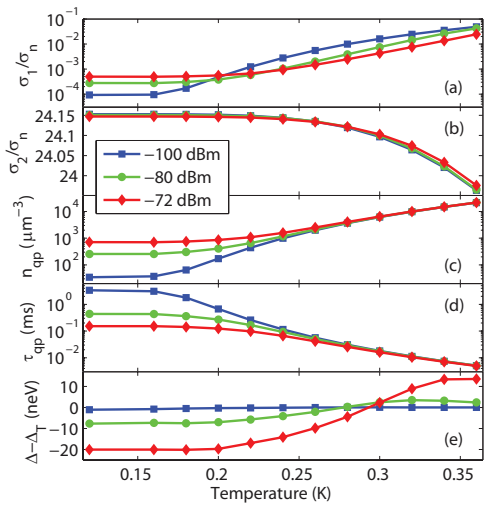}
\caption{\label{fig:simresults}(color online). (a) The real part of the complex conductivity, $\sigma_1$, as a function of temperature, calculated for three microwave readout powers at a frequency of 5.57 GHz. (b) The imaginary part of the conductivity, $\sigma_2$, as a function of temperature. (c) The calculated quasiparticle density as a function of temperature. (d) The quasiparticle recombination lifetime as function of temperature. (e) The difference of the energy gap for the driven distributions ($\Delta$) compared to a thermal distribution ($\Delta_T$). The legend applies to all panels. }

\end{figure}

Having determined the quasiparticle distributions for various readout powers, we can calculate the non-equilibrium conductivity. Fig. \ref{fig:simresults}a and b show $\sigma_1$ and $\sigma_2$, calculated using Eqs. \ref{eq:sigmaone} and \ref{eq:sigmatwo}. For comparison, we plot the quasiparticle density and the quasiparticle recombination time, $\tau_{qp}$, in Fig. \ref{fig:simresults}c and d. At low temperature, we observe that $\sigma_1$ increases and $\sigma_2$ decreases with increasing power, together with an increasing number of quasiparticles (analogous to heating), as described in Ref. \onlinecite{dgoldie2013}. At higher temperatures a counterintuitive effect occurs: $\sigma_1$ decreases (the microwave losses go down) and $\sigma_2$ increases with increasing power, whereas there are still excess quasiparticles being created. This effect cannot be consistently explained with a single effective quasiparticle temperature, but it can be understood from Fig. \ref{fig:distributions}a (at 320 mK). For a thermal $f(E)$, the factor $[f(E)-f(E+\hbar\omega)]$ in Eq. \ref{eq:sigmaone} is larger than for a strongly driven distribution, because of the peaks in the driven distribution with separation $\hbar\omega$. The probability of absorbing a microwave photon is lower for a strongly driven distribution, which decreases $\sigma_1$ and therewith the losses. $\sigma_2$ is only sensitive to quasiparticles at $\Delta<E<\Delta+\hbar\omega$ (Eq. \ref{eq:sigmatwo}). Below 250 mK (see Fig. \ref{fig:simresults}b), the microwave absorption increases the quasiparticle population at $\Delta<E<\Delta+\hbar\omega$, whereas at higher temperatures the population becomes lower due to redistribution. The energy gap, calculated from Eq. \ref{eq:selfconsistent}, is shown in Fig. \ref{fig:simresults}e. Clearly, the non-equilibrium $f(E)$ leads to gap suppression below 0.3 K, and gap enhancement above 0.3 K despite the creation of excess quasiparticles. The additional effect of the non-equilibrium $\Delta$ on the observables is minor, the structure in $f(E)$ dominates.

To connect the calculated $\sigma_1$ and $\sigma_2$ with the experiment, we calculate $Q_i$ and $f_{res}$ through equations for a microstrip geometry \cite{gyassin1995} with the same central strip dimensions as the measured resonator \cite{supinf}. The results are plotted in Figs. \ref{fig:S21Qf}b and c, which shows good agreement with the measurements. In particular, the cross-over temperatures in $Q_i$ and $f_{res}$ are well modelled, as is the temperature dependence of $Q_i$ for both high and low powers. A comparison of Figs. \ref{fig:S21Qf}b and c with Figs. \ref{fig:simresults}a and b shows that $Q_i$ is dominated by $\sigma_1$ and $f_{res}$ by $\sigma_2$, as expected.

\begin{figure}

\includegraphics[width=0.99\columnwidth]{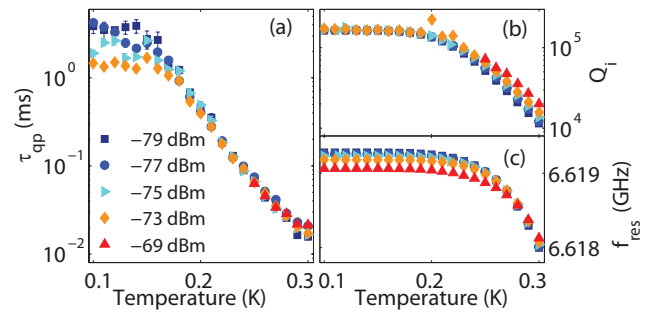}
\caption{\label{fig:lifetimeqf}(color online). (a) The measured quasiparticle recombination time as obtained from a noise measurement for various microwave readout powers measured on sample B. (b) The measured internal quality factor and (c) resonant frequency. The legend applies to all panels.}

\end{figure}

The experimental evidence for the different power dependence of $\tau_{qp}$ and the conductivity is shown in Fig. \ref{fig:lifetimeqf}. These results were measured on sample B \cite{supinf}, on which we performed accurate measurements of $\tau_{qp}$ as reported on in Refs. \onlinecite{pdevisser2011,pdevisser2012b}. Fig. \ref{fig:lifetimeqf}a shows $\tau_{qp}$ as determined from the cross-power spectral density of quasiparticle fluctuations in the amplitude and the phase of the resonator \cite{pdevisser2012b}. Panels b and c show the measured $Q_i$ and $f_{res}$. The power range for this noise measurement is only 10 dB, due to the amplifier noise limit. We focus on $T>200$ mK. $Q_i$ increases with increasing power, consistent with Fig. \ref{fig:S21Qf}b, whereas $\tau_{qp}$ stays constant, as expected from the simulations in Fig. \ref{fig:simresults}d. We thus have a nonlinear conductivity effect due to quasiparticle redistribution, where $Q_i$ increases despite of the creation of excess quasiparticles. This is in contrast with situations in which excess quasiparticles are introduced either on purpose or due to the environment \cite{jgao2008c,jmartinis2009,gcatelani2010,acorcoles2011,rbarends2011,mlenander2011,drainis2012} where $Q_i\propto 1/n_{qp}$, although also in qubits subtleties can occur due to $f(E)$ \cite{jwenner2013}. 

The qualitative agreement between measurements and calculations as apparent from Fig. \ref{fig:S21Qf}b is quite satisfactory. However, the effect of the microwave power on $Q_i$ and $f_{res}$ is less than calculated. Since the uncertainty in the measured $P_{read}$ is less than 2 dB, there should be a parallel dissipation channel. So far we assumed the same $f(E)$ for the groundplane of the resonator and the central strip. Future work may include the calculation of $f(E)$ in the groundplane, which is difficult due to the additional complexity of quasiparticle outdiffusion. A crude approximation, where the groundplane is an impedance with a thermal $f(E)$, in series with the non-equilibrium central strip \cite{rkautz1978}, indicates indeed a reduced non-equilibrium effect of microwave power on $Q_i$ and $f_{res}$. The non-equilibrium f(E) could be measured by combining the resonator experiment with tunnel probes \cite{jwolter1981}.

In closing we emphasize that for the non-equilibrium $f(E)$ to occur (Fig. \ref{fig:distributions}a), quasiparticle-phonon scattering has to be slow compared to $I_{qp}$ and to $\omega$, which is therefore more likely in materials with a low $T_c$, such as Al \cite{skaplan1976}. In addition, redistribution of quasiparticles at low temperatures leads to $n_{qp}\propto\sqrt{P_{abs}}$ \cite{dgoldie2013}, which implies that even in the few microwave photon regime this mechanism leads to excess quasiparticles. 

We would like to thank Y.J.Y. Lankwarden for fabricating the devices. T.M.K. thanks J. Zmuidzinas and A. Vayonakis for discussions on this topic.

%

\clearpage
\onecolumngrid
\setcounter{figure}{0}

\renewcommand{\thefigure}{S\arabic{figure}}

\section*{Supplementary information for: 'Evidence of a nonequilibrium distribution of quasiparticles in the microwave response of a superconducting aluminium resonator'} 

\subsection*{P.J. de Visser, D.J. Goldie, P. Diener, S. Withington, J.J.A. Baselmans and T.M. Klapwijk}


\maketitle

\section{Model input parameters}

An overview of resonator and material parameters is given in Table \ref{tab:deviceparameters}. Measurements reported in the main article in Fig. 1 are measured on Sample A. Data in Fig. 4 is taken on sample B. For $N_0$, the single spin density of states at the Fermi level, we take $1.72\times 10^{4}$ $\mu$eV$^{-1}\mu$m$^{-3}$. $N_0V_{BCS}$ can be found from the measured $T_c$. A microscope picture of a typical resonator is shown in Fig. \ref{fig:circuit}a.

\begingroup
\begin{table}[ht]
\caption{Parameters of the films and resonators for both samples.}\label{tab:deviceparameters}
\begin{tabular}{ccccc}
\hline\hline
Symbol	 & Interpretation			& Sample A 	& Sample B	& Remarks \\

\hline
$t$		&  film thickness			& 60 nm &  40 nm			&  	\\
$T_c$		&  critical temperature		& 1.17 K	&  1.11 K	 &  from DC measurement	\\
$\Delta$		&  energy gap		& 177 $\mu$eV 	&  168 $\mu$eV		&  1.76$k_BT_c$	\\
$\rho_N$			&  low temperature, normal state resistivity	&  0.9 $\mu\Omega$cm	& 0.8 $\mu\Omega$cm	&  from DC measurement	\\
$RRR$		&  residual resistance ratio		&  4.5		& 5.2			&  from DC measurement	\\
$Q_c$		&  coupling quality factor		& 20100	&  54285		&  measured	\\
$C$		&  coupling capacitance			& 7.5 fF	 &  3.7 fF	&  Calculated from $Q_c$	\\
$w$	&  width of central strip of resonator			&  3 $\mu$m			& 3 $\mu$m	&  	\\
$s$		&  gap width of CPW		&  2 $\mu$m			& 2 $\mu$m 		&    \\		
$l$	 	&  length of resonator		& 9.84 mm	&  8.33 mm				&  	\\
$V$		&  central strip volume 	& 1770 $\mu$m$^3$ &	 1000	$\mu$m$^3$	&$V=wtl$\\
$f_0$	  	&  measured resonant frequency	& 5.28 GHz		&  6.62 GHz		&  	\\
$\tau_0$			&  electron-phonon time		& 440 ns		&  440 ns		&  measured in Ref. \onlinecite{pdevisser2011s}	\\
$\tau_{0,ph}$	&  phonon time		&  0.26 ns		& 0.26 ns	&  Ref. \onlinecite{dgoldie2013s}	\\
$\tau_{esc}$	&  phonon escape time 		& 0.17 ns	&  0.11 ns		&  calculated for Al on sapphire from Ref. \onlinecite{skaplan1979s}	\\
$\tau_{pb}$		&  pair breaking time	&  0.28 ns		& 0.28 ns		&  from Kaplan \cite{skaplan1976s}	\\	

\hline\hline
\end{tabular}
\end{table}
\endgroup

\begingroup
\begin{table}[ht]
\caption{Parameters of the modelled microstrip geometry}\label{tab:microstrip}
\begin{tabular}{ccccc}
\hline\hline
Symbol	 & Interpretation			& Value \\

\hline
$t$		&  film thickness					& 60 nm 	\\
$Q_c$		&  coupling quality factor		& 20100		\\
$C$		&  coupling capacitance			& 7.5 fF	 	\\
$w$		&  width of central strip of resonator			&  3 $\mu$m		\\
$s$		&  thickness of the dielectric		&  1 $\mu$m		  \\		
$l$	 	&  length of resonator			& 9.84 mm	\\
$V$		&  central strip volume 		& 1770 $\mu$m$^3$ \\
$f_0$	  	&  modeled resonant frequency	& 5.57 GHz	\\
\hline\hline
\end{tabular}
\end{table}
\endgroup

\section{Absorbed microwave power in the quasiparticle system}

\begin{figure}[ht]

\includegraphics[width=0.85\columnwidth]{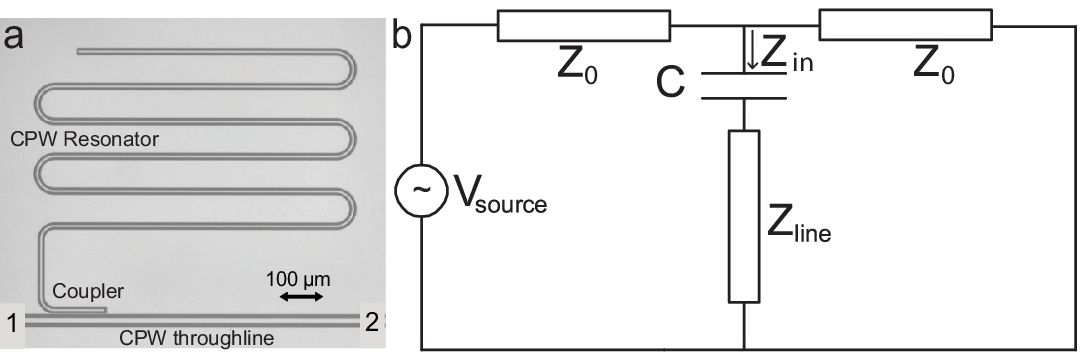}
\caption{\label{fig:circuit} (a) A microscope picture of one of the halfwave microwave resonators. The resonator is shortened two times in length for visibility. The coplanar waveguide (CPW) through transmission line is used for readout. The transmission from port 1 to port 2 is measured. (b) The microwave circuit considered for calculating the absorbed microwave power. }

\end{figure}

For the modelling of the effect of absorbed microwave powers we need to derive the absorbed power in the quasiparticle system, $P_{abs}$, in a way consistent with how the quality factors and resonant frequency are determined. To that end we consider the circuit in Fig. \ref{fig:circuit}b. The resonator is a half-wave resonator with two open ends as used in the experiment, which is coupled with a coupling capacitance $C$ to a through transmission line with a characteristic impedance $Z_0$ (typically 50 $\Omega$). The geometry can be seen as a shunt resonator (it shunts the through transmission line). Similar expressions as below are reported for different geometries \cite{bmazinphd,rbarendsphd,jzmuidzinas2012s}, but we derive them here explicitly for the halfwave geometry, because $P_{abs}$ is a crucial parameter to connect the experiments and the simulations. The main results are Eq. \ref{eq:S21magsquared}, relating $S_{21}$ to the quality factors, and Eq. \ref{eq:Pdissqp}, which relates $P_{abs}$ to the readout power and the quality factors.

The input impedance of a halfwave transmission line resonator with two open ends is given by \cite{pozar}
\begin{equation}\label{eq:halfwaveimpedance}
Z_{line} = Z_0\coth(\alpha+j\beta)l = Z_0\frac{1+j\tan\beta l\tanh \alpha l}{\tanh \alpha l + j\tan\beta l},
\end{equation}
where $l$ is the length of the line. $\alpha$ and $\beta$ are the real and imaginary parts of the propagation constant. At a wavelength of twice the length, $l=\lambda/2$, close to the resonant frequency $\omega_{1/2}$ (that is: the resonant frequency of the half wavelength line without coupler), with $\omega=\omega_{1/2}+\Delta\omega$, 
\begin{eqnarray}
\beta l = \pi + \frac{\pi\Delta\omega}{\omega_{1/2}},\\
\tan\beta l = \tan\frac{\pi\Delta\omega}{\omega_0}\approx\frac{\Delta\omega\pi}{\omega_{1/2}}.
\end{eqnarray}
Also using that $\tanh\alpha l\approx\alpha l$ and $j\tan\beta l\tanh \alpha l\approx\alpha l\frac{\pi\Delta\omega}{\omega_{1/2}}$, we can rewrite Eq. \ref{eq:halfwaveimpedance} as
\begin{equation}
Z_{line}=\frac{Z_0}{\alpha l + j\frac{\Delta\omega\pi}{\omega_{1/2}}}.
\end{equation}
We now define the internal quality factor of this halfwave resonator as $Q_i=\frac{\alpha}{2\beta}$, which leads to $\alpha l = \frac{\beta l}{2 Q_i}=\frac{\pi}{2Q_i}(1+\frac{\Delta\omega}{\omega_{1/2}})$. Therefore we get
\begin{equation}
	Z_{line} = Z_0\frac{1}{\frac{\pi}{2Q_i}(1+\frac{\Delta\omega}{\omega_{1/2}})+j\frac{\pi\Delta\omega}{\omega_{1/2}}}=Z_0\frac{2Q_i/\pi}{(1+\frac{\Delta\omega}{\omega_{1/2}})+j2Q_i\frac{\pi\Delta\omega}{\omega_{1/2}}}=Z_0\frac{\frac{2Q_i}{\pi}-4j\frac{Q_i^2}{\pi}\frac{\Delta\omega}{\omega_{1/2}}}{1+4Q_i^2\left(\frac{\Delta\omega}{\omega_{1/2}}\right)^2},
\end{equation}
where for the last step we used $\frac{\Delta\omega}{\omega_{1/2}}<<1$.

We add a capacitive coupler on one side of the halfwave resonator with a capacitance $C$. Now the impedance looking into the capacitance is given by
\begin{equation}\label{eq:impedancelineandcoupler}
Z_{in} = Z_{line}+\frac{1}{j\omega C} = Z_0\left(\frac{\frac{2Q_i}{\pi}-4j\frac{Q_i^2}{\pi}\frac{\Delta\omega}{\omega_{1/2}}}{1+4Q_i^2\left(\frac{\Delta\omega}{\omega_{1/2}}\right)^2} - \frac{j}{\omega C Z_0} \right)=Z_0\frac{\frac{2Qi}{\pi}-4j\frac{Q_i^2}{\pi}\frac{\Delta\omega}{\omega_{1/2}}-\frac{j}{\omega C Z_0}\left( 1+4Q_i^2\left(\frac{\Delta\omega}{\omega_{1/2}}\right)^2 \right)}{1+4Q_i^2\left(\frac{\Delta\omega}{\omega_{1/2}}\right)^2}
\end{equation}
The added capacitance will cause the resonant frequency of the line and coupler, $\omega_0$, to deviate from the resonant frequency of the line only, $\omega_{1/2}$. The new criterion for resonance will be that the imaginary part of Eq. \ref{eq:impedancelineandcoupler} vanishes, $\Im (Z_{in})=0$. That requires that
\begin{equation}
\frac{4 Q_i^2}{\omega CZ_0}\left(\frac{\Delta \omega}{\omega_{1/2}} \right)^2 + \frac{4 Q_i^2}{\pi}\left(\frac{\Delta \omega}{\omega_{1/2}} \right)+\frac{1}{\omega CZ_0} = 0,
\end{equation}
the solutions of which are
\begin{equation}\label{eq:resonancesolution}
\frac{\Delta \omega}{\omega_{1/2}} = \frac{\omega CZ_0}{2Q_i}\left(-\frac{Q_i}{\pi} \pm \sqrt{\frac{Q_i^2}{\pi}-\frac{1}{\omega^2C^2Z_0^2}} \right)
\approx \frac{\omega C Z_0}{2Q_i}\left(-\frac{Q_i}{\pi} \pm \frac{Q_i}{\pi} \right)
= -\frac{\omega C Z_0}{\pi}\vee 0.
\end{equation}
The approximation here requires that $\frac{1}{\omega^2C^2Z_0^2}<<Q_i^2$, which holds for all practical cases. We take the first solution, which lowers the resonant frequency and makes the line slightly inductive. For that solution
\begin{equation}\label{eq:realimpedance}
\Re(Z_{in}) = \frac{2Z_0Q_i/\pi}{1+4Q_i^2\left(\frac{\Delta\omega}{\omega_{1/2}}\right)^2}
=\frac{2Z_0Q_i/\pi}{1+4Q_i^2\omega^2Z_0^2C^2/\pi^2}
\approx \frac{Z_0\pi}{2Q_i\omega^2C^2Z_0^2}.
\end{equation}

We will now first derive the coupling quality factor $Q_c$. The energy stored in the line at resonance is $\frac{1}{2}C_lV_{line}^2$, where $C_l$ is the (total distributed) capacitance of the halfwave line and $V_{line}$ the voltage over the line. If we divide the voltage $V_L$ over the coupler and the halfwave line, we get $V_{line} = \frac{Z_{in}}{Z_{line}}V_L$. The current is given by $I=V_L/Z_{in} = V_{line}/Z_{line}$. The dissipated power in the throughline is 
\begin{equation}
P_{diss}^c = |I|^2\frac{Z_0}{2} = \frac{V_{line}^2}{|Z_{line}|^2}\frac{Z_0}{2}.
\end{equation}
The superscript $c$ is to denote that it is the power that is lost through the coupler.
At resonance, $\omega=\omega_0$, we get from above that $|Z_{line}|=\frac{1}{\omega C}$. Furthermore we can write the capacitance of the line to be $C_l=\frac{L}{v_{ph}Z_0}$, where $v_{ph}$ is the phase velocity and we assume again that the characteristic impedance is $Z_0$. At the first resonance of the circuit we can write $\omega C_l=\frac{\omega L}{v_{ph} Z_0} = \frac{2\pi L}{\lambda Z_0} = \frac{\pi}{Z_0}$, where $\lambda$ is the wavelength. Now
\begin{equation}\label{eq:Qcsecond}
Q_c = \frac{\omega E}{P_{diss}^c} = \frac{\omega\frac{1}{2}C_lV_{line}^2}{\frac{V_{line}^2Z_0/2}{|Z_{line}|^2}} = \frac{\omega C_l|Z_{line}|^2}{Z_0} = \frac{\pi}{\omega^2 C^2 Z_0^2}.
\end{equation}

We can now rewrite our equations for small deviations with respect to the new resonant frequency (of the halfwave line \emph{and} coupler) $\omega_0$ using $\Delta\omega/\omega_{1/2} = \Delta\omega/\omega_0 - \sqrt{1/\pi Q_c}$. If we substitute that in Eq. \ref{eq:impedancelineandcoupler} we get
\begin{equation}
\frac{Z_{in}}{Z_0} = \frac{2Q_i/\pi}{1+2jQ_i\left(\frac{\Delta\omega}{\omega_0}-j\sqrt{\frac{1}{\pi Q_c}} \right)}-j\sqrt{\frac{Q_c}{\pi}}
= \frac{2Q_i\sqrt{\frac{Q_c}{\pi}}\frac{\Delta\omega}{\omega_0}-j\sqrt{\frac{1}{\pi Q_c}}}{1+2jQ_i\left(\frac{\Delta\omega}{\omega_0}-\sqrt{\frac{1}{\pi Q_c}} \right)}
\end{equation}

On resonance, $\frac{\Delta\omega}{\omega_0}=0$ and
\begin{equation}\label{eq:Zinatresonance}
\frac{Z_{in}}{Z_0}= \frac{2Q_i\sqrt{\frac{Q_c}{\pi}}-j\sqrt{\frac{1}{\pi Q_c}}}{1-2jQ_i\sqrt{\frac{1}{\pi Q_c}}}=\frac{-j\sqrt{\frac{1}{\pi Q_c}}+2\frac{Q_i}{\pi}}{1+4\frac{Q_i^2}{\pi Q_c}} \approx \frac{Q_c}{2Q_i},
\end{equation}
which shows that at critical coupling ($Q_i=Q_c$) and at resonance $Z_{in}=Z_0/2$.

The scattering parameter $S_{21}$, is for the circuit in Fig. \ref{fig:circuit}b given by
\begin{equation}
S_{21}=\frac{2}{2+\frac{Z_0}{Z}} = \frac{2Q_i\frac{\Delta\omega}{\omega_0}-j}{\frac{1}{2}\sqrt{\frac{\pi}{Q_c}}+ 2Q_i\frac{\Delta\omega}{\omega_0}-j+jQ_i\sqrt{\frac{\pi}{Q_c}}\left( \frac{\Delta\omega}{\omega_0}-\sqrt{\frac{1}{\pi Q_c}} \right)}
=\frac{Q/Q_i+2jQ\frac{\Delta\omega}{\omega_0}}{1+2jQ\frac{\Delta\omega}{\omega_0}},
\end{equation}
where $Q$ is the loaded quality factor: $Q=\frac{Q_iQ_c}{Q_i+Q_c}$. In practice we usually fit $|S_{21}|^2$, which can be readily derived:
\begin{equation}\label{eq:S21magsquared}
|S_{21}|^2 = \frac{S_{21,min}^2+4Q^2\left(\frac{\Delta\omega}{\omega_0} \right)^2}{1+4Q^2\left(\frac{\Delta\omega}{\omega_0} \right)^2} = 1 + \frac{S_{21,min}^2-1}{1+4Q^2\left(\frac{\Delta\omega}{\omega_0} \right)^2},
\end{equation}
where $S_{21,min}$ is the transmission on resonance, $S_{21,min}=\frac{Q}{Q_i}$.
This is a result that is very commonly used to fit resonance curves and is used in this work to extract the quality factors and the resonant frequency from the resonance curves.

In Fig. \ref{fig:circuit}b we consider the load of this circuit to be $Z_{in}$, through which a current $I_L$ flows and over which a voltage $V_L$ is applied. Therefore 
\begin{equation}
I_L = \frac{V_{source}}{2Z_{in}+Z_0} = \frac{V_{source}/Z_{in}}{2+\frac{Z_0}{Z_{in}}}.
\end{equation}
The dissipated power in the load is now given by
\begin{equation}\label{eq:Pdiss}
P_{abs} = |I_L|^2\Re(Z_{in}) = \frac{V_{source}^2\Re(Z_{in})}{|2Z_{in}+Z_0|^2} = \frac{4Z_0P_{read}\Re(Z_{in})}{|2Z_{in}+Z_0|^2},
\end{equation}
where we take $P_{read}=\frac{V_{source}^2}{4Z_0}$ to be the readout power on the throughline. 

We now focus on the dissipated power at resonance. From Eq. \ref{eq:Zinatresonance}, $\frac{Z_{in}}{Z_0} = \frac{Q_c}{2Q_i}$. Therefore
\begin{equation}\label{eq:Pdissatresonance}
P_{abs} = \frac{2Z_0^2P_{read}\frac{Q_c}{Q_i}}{Z_0^2\left(\frac{Q_c}{Q_i}+1\right)^2} = \frac{P_{read}}{2}\frac{4Q^2}{Q_iQ_c}.
\end{equation}
At critical coupling $Q_i=Q_c$, $P_{abs}=P_{read}/2$.

So far we have assumed that there is only one source of dissipation associated with $Q_i$. We can refine this assumption by adding an additional source of loss. It is always possible to write this loss as a quality factor, say $Q_i^*$. We define $Q_i^{qp}$ as the quality factor due to dissipation in the quasiparticle system. Then
\begin{equation}
\frac{1}{Q} = \frac{1}{Q_c} + \frac{1}{Q_i^*} + \frac{1}{Q_i^{qp}} = \frac{1}{Q_c} + \frac{1}{Q_i},
\end{equation}
and therefore
\begin{equation}\label{eq:Pdissqp}
P_{abs}^{qp} = P_{abs} \frac{Q_i}{Q_i^{qp}} = \frac{P_{read}}{2}\frac{4Q^2}{Q_iQ_c}\frac{Q_i}{Q_i^{qp}}
\end{equation}
where we assumed to be on resonance (Eq. \ref{eq:Pdissatresonance}). Also this equation has been reported \cite{jzmuidzinas2012s}, but it is important to notice that with this derivation we have shown that one can consistently derive $P_{abs}$ from parameters that are obtained from a fit to the resonance curve as given by Eq. \ref{eq:S21magsquared}.

\begin{figure}

\includegraphics[width=0.4\columnwidth]{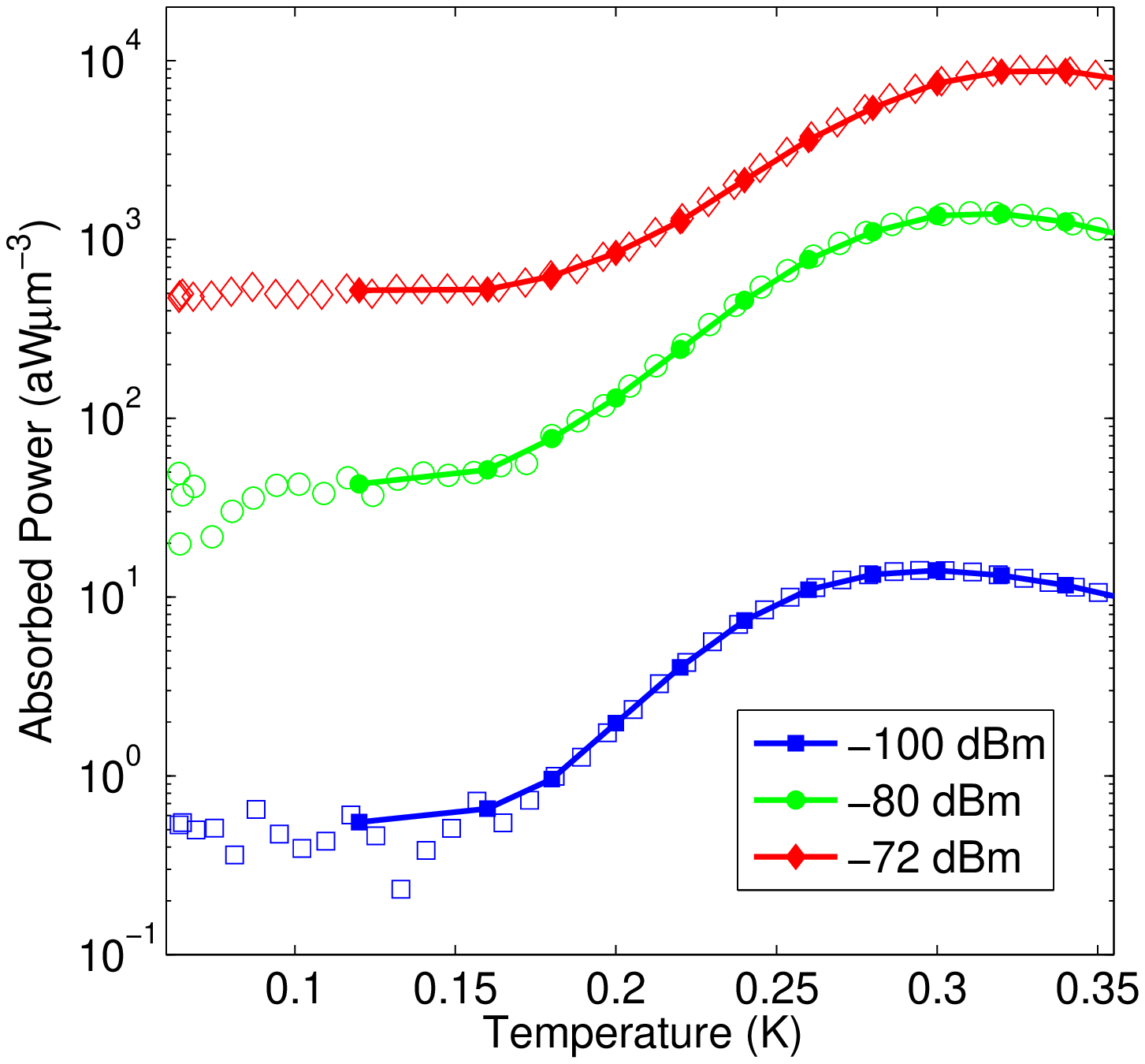}
\caption{\label{fig:Pabs} The absorbed microwave power per unit volume as a function of temperature for three different microwave readout powers. The open symbols are derived from the measurements. The filled symbols with lines show that in the simulations we have modeled the same absorbed powers as derived from the measurements. }

\end{figure}

$P_{abs}$ as determined through this procedure on sample A is shown as a function of temperature for several microwave powers in Fig. \ref{fig:Pabs}. We observe that the absorbed power not only scales with $P_{read}$ but is also strongly temperature dependent. This temperature dependence can be understood through Eq. \ref{eq:Pdissatresonance} by considering that $Q_i>>Q_c$ at 100 mK and $Q_i\approx Q_c$ close to 300 mK. 

To compare parameters that follow from the calculated quasiparticle distribution functions we convert $\sigma_1$ and $\sigma_2$ into a resonant frequency and an internal quality factor, using the expressions for a microstrip geometry as can be found in Ref. \onlinecite{gyassin1995s}. These equations allow to calculate the complex propagation constant of the line, including the capacitance, the geometric and kinetic inductance and the losses and therewith the impedance of the line. Equations for coplanar waveguide geometries usually give separate approximations for the attenuation and the kinetic inductance\cite{jclem2013}, or only one of the two \cite{cholloway1995}. The central strip dimensions and the film thickness have been kept the same as for the measured coplanar waveguide geometry. Since we are mainly interested here in the properties of the superconductor, these geometries are comparable.

%


\begin{thebibliography}{42}%
\makeatletter
\providecommand \@ifxundefined [1]{%
 \@ifx{#1\undefined}
}%
\providecommand \@ifnum [1]{%
 \ifnum #1\expandafter \@firstoftwo
 \else \expandafter \@secondoftwo
 \fi
}%
\providecommand \@ifx [1]{%
 \ifx #1\expandafter \@firstoftwo
 \else \expandafter \@secondoftwo
 \fi
}%
\providecommand \natexlab [1]{#1}%
\providecommand \enquote  [1]{``#1''}%
\providecommand \bibnamefont  [1]{#1}%
\providecommand \bibfnamefont [1]{#1}%
\providecommand \citenamefont [1]{#1}%
\providecommand \href@noop [0]{\@secondoftwo}%
\providecommand \href [0]{\begingroup \@sanitize@url \@href}%
\providecommand \@href[1]{\@@startlink{#1}\@@href}%
\providecommand \@@href[1]{\endgroup#1\@@endlink}%
\providecommand \@sanitize@url [0]{\catcode `\\12\catcode `\$12\catcode
  `\&12\catcode `\#12\catcode `\^12\catcode `\_12\catcode `\%12\relax}%
\providecommand \@@startlink[1]{}%
\providecommand \@@endlink[0]{}%
\providecommand \url  [0]{\begingroup\@sanitize@url \@url }%
\providecommand \@url [1]{\endgroup\@href {#1}{\urlprefix }}%
\providecommand \urlprefix  [0]{URL }%
\providecommand \Eprint [0]{\href }%
\providecommand \doibase [0]{http://dx.doi.org/}%
\providecommand \selectlanguage [0]{\@gobble}%
\providecommand \bibinfo  [0]{\@secondoftwo}%
\providecommand \bibfield  [0]{\@secondoftwo}%
\providecommand \translation [1]{[#1]}%
\providecommand \BibitemOpen [0]{}%
\providecommand \bibitemStop [0]{}%
\providecommand \bibitemNoStop [0]{.\EOS\space}%
\providecommand \EOS [0]{\spacefactor3000\relax}%
\providecommand \BibitemShut  [1]{\csname bibitem#1\endcsname}%
\let\auto@bib@innerbib\@empty
\bibitem [{Note1()}]{Note1}%
  \BibitemOpen
  \bibinfo {note} {This pair breaking mechanism is different from direct Cooper
  pair breaking by e.g. photons.}\BibitemShut {Stop}%
\bibitem [{\citenamefont {Yip}\ and\ \citenamefont {Sauls}(1992)}]{syip1992}%
  \BibitemOpen
  \bibfield  {author} {\bibinfo {author} {\bibfnamefont {S.~K.}\ \bibnamefont
  {Yip}}\ and\ \bibinfo {author} {\bibfnamefont {J.~A.}\ \bibnamefont
  {Sauls}},\ }\href {\doibase 10.1103/PhysRevLett.69.2264} {\bibfield
  {journal} {\bibinfo  {journal} {Phys. Rev. Lett.}\ }\textbf {\bibinfo
  {volume} {69}},\ \bibinfo {pages} {2264} (\bibinfo {year}
  {1992})}\BibitemShut {NoStop}%
\bibitem [{\citenamefont {Chin}\ \emph {et~al.}(1992)\citenamefont {Chin},
  \citenamefont {Oates}, \citenamefont {Dresselhaus},\ and\ \citenamefont
  {Dresselhaus}}]{cchin1992}%
  \BibitemOpen
  \bibfield  {author} {\bibinfo {author} {\bibfnamefont {C.~C.}\ \bibnamefont
  {Chin}}, \bibinfo {author} {\bibfnamefont {D.~E.}\ \bibnamefont {Oates}},
  \bibinfo {author} {\bibfnamefont {G.}~\bibnamefont {Dresselhaus}}, \ and\
  \bibinfo {author} {\bibfnamefont {M.~S.}\ \bibnamefont {Dresselhaus}},\
  }\href@noop {} {\bibfield  {journal} {\bibinfo  {journal} {Phys. Rev. B}\
  }\textbf {\bibinfo {volume} {45}},\ \bibinfo {pages} {4788} (\bibinfo {year}
  {1992})}\BibitemShut {NoStop}%
\bibitem [{\citenamefont {Eom}\ \emph {et~al.}(2012)\citenamefont {Eom},
  \citenamefont {Day}, \citenamefont {LeDuc},\ and\ \citenamefont
  {Zmuidzinas}}]{beom2012}%
  \BibitemOpen
  \bibfield  {author} {\bibinfo {author} {\bibfnamefont {B.~H.}\ \bibnamefont
  {Eom}}, \bibinfo {author} {\bibfnamefont {P.~K.}\ \bibnamefont {Day}},
  \bibinfo {author} {\bibfnamefont {H.~G.}\ \bibnamefont {LeDuc}}, \ and\
  \bibinfo {author} {\bibfnamefont {J.}~\bibnamefont {Zmuidzinas}},\
  }\href@noop {} {\bibfield  {journal} {\bibinfo  {journal} {Nature Phys.}\
  }\textbf {\bibinfo {volume} {8}},\ \bibinfo {pages} {623} (\bibinfo {year}
  {2012})}\BibitemShut {NoStop}%
\bibitem [{\citenamefont {Chang}\ and\ \citenamefont
  {Scalapino}(1977)}]{jchang1977}%
  \BibitemOpen
  \bibfield  {author} {\bibinfo {author} {\bibfnamefont {J.-J.}\ \bibnamefont
  {Chang}}\ and\ \bibinfo {author} {\bibfnamefont {D.~J.}\ \bibnamefont
  {Scalapino}},\ }\href@noop {} {\bibfield  {journal} {\bibinfo  {journal}
  {Phys. Rev. B}\ }\textbf {\bibinfo {volume} {15}},\ \bibinfo {pages} {2651}
  (\bibinfo {year} {1977})}\BibitemShut {NoStop}%
\bibitem [{\citenamefont {Wolter}\ and\ \citenamefont
  {Horstman}(1981)}]{jwolter1981}%
  \BibitemOpen
  \bibfield  {author} {\bibinfo {author} {\bibfnamefont {J.}~\bibnamefont
  {Wolter}}\ and\ \bibinfo {author} {\bibfnamefont {R.~E.}\ \bibnamefont
  {Horstman}},\ }\href@noop {} {\bibfield  {journal} {\bibinfo  {journal}
  {Phys. Lett.}\ }\textbf {\bibinfo {volume} {86A}},\ \bibinfo {pages} {185}
  (\bibinfo {year} {1981})}\BibitemShut {NoStop}%
\bibitem [{\citenamefont {Eliashberg}(1970)}]{geliashberg1970}%
  \BibitemOpen
  \bibfield  {author} {\bibinfo {author} {\bibfnamefont {G.~M.}\ \bibnamefont
  {Eliashberg}},\ }\href@noop {} {\bibfield  {journal} {\bibinfo  {journal}
  {JETP Lett.}\ }\textbf {\bibinfo {volume} {11}},\ \bibinfo {pages} {114}
  (\bibinfo {year} {1970})}\BibitemShut {NoStop}%
\bibitem [{\citenamefont {Ivlev}\ \emph {et~al.}(1973)\citenamefont {Ivlev},
  \citenamefont {Lisitsyn},\ and\ \citenamefont {Eliashberg}}]{bivlev1973}%
  \BibitemOpen
  \bibfield  {author} {\bibinfo {author} {\bibfnamefont {B.~I.}\ \bibnamefont
  {Ivlev}}, \bibinfo {author} {\bibfnamefont {S.~G.}\ \bibnamefont {Lisitsyn}},
  \ and\ \bibinfo {author} {\bibfnamefont {G.~M.}\ \bibnamefont {Eliashberg}},\
  }\href@noop {} {\bibfield  {journal} {\bibinfo  {journal} {J. Low Temp.
  Phys.}\ }\textbf {\bibinfo {volume} {10}},\ \bibinfo {pages} {449} (\bibinfo
  {year} {1973})}\BibitemShut {NoStop}%
\bibitem [{\citenamefont {Klapwijk}\ and\ \citenamefont
  {Mooij}(1976)}]{tklapwijk1976}%
  \BibitemOpen
  \bibfield  {author} {\bibinfo {author} {\bibfnamefont {T.~M.}\ \bibnamefont
  {Klapwijk}}\ and\ \bibinfo {author} {\bibfnamefont {J.~E.}\ \bibnamefont
  {Mooij}},\ }\href@noop {} {\bibfield  {journal} {\bibinfo  {journal} {Physica
  B+C}\ }\textbf {\bibinfo {volume} {81}},\ \bibinfo {pages} {132} (\bibinfo
  {year} {1976})}\BibitemShut {NoStop}%
\bibitem [{\citenamefont {Klapwijk}\ \emph {et~al.}(1977)\citenamefont
  {Klapwijk}, \citenamefont {van~den Bergh},\ and\ \citenamefont
  {Mooij}}]{tklapwijk1977}%
  \BibitemOpen
  \bibfield  {author} {\bibinfo {author} {\bibfnamefont {T.~M.}\ \bibnamefont
  {Klapwijk}}, \bibinfo {author} {\bibfnamefont {J.~N.}\ \bibnamefont {van~den
  Bergh}}, \ and\ \bibinfo {author} {\bibfnamefont {J.~E.}\ \bibnamefont
  {Mooij}},\ }\href@noop {} {\bibfield  {journal} {\bibinfo  {journal} {J. Low
  Temp. Phys.}\ }\textbf {\bibinfo {volume} {26}},\ \bibinfo {pages} {385}
  (\bibinfo {year} {1977})}\BibitemShut {NoStop}%
\bibitem [{\citenamefont {Horstman}\ and\ \citenamefont
  {Wolter}(1981)}]{rhorstman1981}%
  \BibitemOpen
  \bibfield  {author} {\bibinfo {author} {\bibfnamefont {R.~E.}\ \bibnamefont
  {Horstman}}\ and\ \bibinfo {author} {\bibfnamefont {J.}~\bibnamefont
  {Wolter}},\ }\href@noop {} {\bibfield  {journal} {\bibinfo  {journal} {Phys.
  Lett.}\ }\textbf {\bibinfo {volume} {82A}},\ \bibinfo {pages} {43} (\bibinfo
  {year} {1981})}\BibitemShut {NoStop}%
\bibitem [{\citenamefont {Sridhar}\ and\ \citenamefont
  {Mercereau}(1980)}]{ssridhar1980}%
  \BibitemOpen
  \bibfield  {author} {\bibinfo {author} {\bibfnamefont {S.}~\bibnamefont
  {Sridhar}}\ and\ \bibinfo {author} {\bibfnamefont {J.~E.}\ \bibnamefont
  {Mercereau}},\ }\href@noop {} {\bibfield  {journal} {\bibinfo  {journal}
  {Phys. Lett. A}\ }\textbf {\bibinfo {volume} {75}},\ \bibinfo {pages} {392}
  (\bibinfo {year} {1980})}\BibitemShut {NoStop}%
\bibitem [{\citenamefont {Goldie}\ and\ \citenamefont
  {Withington}(2013)}]{dgoldie2013}%
  \BibitemOpen
  \bibfield  {author} {\bibinfo {author} {\bibfnamefont {D.~J.}\ \bibnamefont
  {Goldie}}\ and\ \bibinfo {author} {\bibfnamefont {S.}~\bibnamefont
  {Withington}},\ }\href@noop {} {\bibfield  {journal} {\bibinfo  {journal}
  {Supercond. Sci. Technol.}\ }\textbf {\bibinfo {volume} {26}},\ \bibinfo
  {pages} {015004} (\bibinfo {year} {2013})}\BibitemShut {NoStop}%
\bibitem [{\citenamefont {de~Visser}\ \emph {et~al.}(2012)\citenamefont
  {de~Visser}, \citenamefont {Baselmans}, \citenamefont {Yates}, \citenamefont
  {Diener}, \citenamefont {Endo},\ and\ \citenamefont
  {Klapwijk}}]{pdevisser2012b}%
  \BibitemOpen
  \bibfield  {author} {\bibinfo {author} {\bibfnamefont {P.~J.}\ \bibnamefont
  {de~Visser}}, \bibinfo {author} {\bibfnamefont {J.~J.~A.}\ \bibnamefont
  {Baselmans}}, \bibinfo {author} {\bibfnamefont {S.~J.~C.}\ \bibnamefont
  {Yates}}, \bibinfo {author} {\bibfnamefont {P.}~\bibnamefont {Diener}},
  \bibinfo {author} {\bibfnamefont {A.}~\bibnamefont {Endo}}, \ and\ \bibinfo
  {author} {\bibfnamefont {T.~M.}\ \bibnamefont {Klapwijk}},\ }\href@noop {}
  {\bibfield  {journal} {\bibinfo  {journal} {Appl. Phys. Lett.}\ }\textbf
  {\bibinfo {volume} {100}},\ \bibinfo {pages} {162601} (\bibinfo {year}
  {2012})}\BibitemShut {NoStop}%
\bibitem [{\citenamefont {Day}\ \emph {et~al.}(2003)\citenamefont {Day},
  \citenamefont {LeDuc}, \citenamefont {Mazin}, \citenamefont {Vayonakis},\
  and\ \citenamefont {Zmuidzinas}}]{pday2003}%
  \BibitemOpen
  \bibfield  {author} {\bibinfo {author} {\bibfnamefont {P.~K.}\ \bibnamefont
  {Day}}, \bibinfo {author} {\bibfnamefont {H.~G.}\ \bibnamefont {LeDuc}},
  \bibinfo {author} {\bibfnamefont {B.~A.}\ \bibnamefont {Mazin}}, \bibinfo
  {author} {\bibfnamefont {A.}~\bibnamefont {Vayonakis}}, \ and\ \bibinfo
  {author} {\bibfnamefont {J.}~\bibnamefont {Zmuidzinas}},\ }\href@noop {}
  {\bibfield  {journal} {\bibinfo  {journal} {Nature}\ }\textbf {\bibinfo
  {volume} {425}},\ \bibinfo {pages} {817} (\bibinfo {year}
  {2003})}\BibitemShut {NoStop}%
\bibitem [{\citenamefont {Zmuidzinas}(2012)}]{jzmuidzinas2012}%
  \BibitemOpen
  \bibfield  {author} {\bibinfo {author} {\bibfnamefont {J.}~\bibnamefont
  {Zmuidzinas}},\ }\href@noop {} {\bibfield  {journal} {\bibinfo  {journal}
  {Ann. Rev. Condens. Matter Phys.}\ }\textbf {\bibinfo {volume} {3}},\
  \bibinfo {pages} {169} (\bibinfo {year} {2012})}\BibitemShut {NoStop}%
\bibitem [{\citenamefont {Chiodi}\ \emph {et~al.}(2009)\citenamefont {Chiodi},
  \citenamefont {Aprili},\ and\ \citenamefont {Reulet}}]{fchiodi2009}%
  \BibitemOpen
  \bibfield  {author} {\bibinfo {author} {\bibfnamefont {F.}~\bibnamefont
  {Chiodi}}, \bibinfo {author} {\bibfnamefont {M.}~\bibnamefont {Aprili}}, \
  and\ \bibinfo {author} {\bibfnamefont {B.}~\bibnamefont {Reulet}},\
  }\href@noop {} {\bibfield  {journal} {\bibinfo  {journal} {Phys. Rev. Lett.}\
  }\textbf {\bibinfo {volume} {103}},\ \bibinfo {pages} {177002} (\bibinfo
  {year} {2009})}\BibitemShut {NoStop}%
\bibitem [{\citenamefont {Virtanen}\ \emph {et~al.}(2010)\citenamefont
  {Virtanen}, \citenamefont {Heikkil\"{a}}, \citenamefont {Bergeret},\ and\
  \citenamefont {Cuevas}}]{pvirtanen2010}%
  \BibitemOpen
  \bibfield  {author} {\bibinfo {author} {\bibfnamefont {P.}~\bibnamefont
  {Virtanen}}, \bibinfo {author} {\bibfnamefont {T.~T.}\ \bibnamefont
  {Heikkil\"{a}}}, \bibinfo {author} {\bibfnamefont {F.~S.}\ \bibnamefont
  {Bergeret}}, \ and\ \bibinfo {author} {\bibfnamefont {J.~C.}\ \bibnamefont
  {Cuevas}},\ }\href@noop {} {\bibfield  {journal} {\bibinfo  {journal} {Phys.
  Rev. Lett.}\ }\textbf {\bibinfo {volume} {104}},\ \bibinfo {pages} {247003}
  (\bibinfo {year} {2010})}\BibitemShut {NoStop}%
\bibitem [{\citenamefont {Beck}\ \emph {et~al.}(2013)\citenamefont {Beck},
  \citenamefont {Rousseau}, \citenamefont {Klammer}, \citenamefont {Leiderer},
  \citenamefont {Mittendorff}, \citenamefont {Winnerl}, \citenamefont {Helm},
  \citenamefont {Gol'tsman},\ and\ \citenamefont {Demsar}}]{mbeck2013}%
  \BibitemOpen
  \bibfield  {author} {\bibinfo {author} {\bibfnamefont {M.}~\bibnamefont
  {Beck}}, \bibinfo {author} {\bibfnamefont {I.}~\bibnamefont {Rousseau}},
  \bibinfo {author} {\bibfnamefont {M.}~\bibnamefont {Klammer}}, \bibinfo
  {author} {\bibfnamefont {P.}~\bibnamefont {Leiderer}}, \bibinfo {author}
  {\bibfnamefont {M.}~\bibnamefont {Mittendorff}}, \bibinfo {author}
  {\bibfnamefont {S.}~\bibnamefont {Winnerl}}, \bibinfo {author} {\bibfnamefont
  {M.}~\bibnamefont {Helm}}, \bibinfo {author} {\bibfnamefont {G.~N.}\
  \bibnamefont {Gol'tsman}}, \ and\ \bibinfo {author} {\bibfnamefont
  {J.}~\bibnamefont {Demsar}},\ }\href {\doibase
  10.1103/PhysRevLett.110.267003} {\bibfield  {journal} {\bibinfo  {journal}
  {Phys. Rev. Lett.}\ }\textbf {\bibinfo {volume} {110}},\ \bibinfo {pages}
  {267003} (\bibinfo {year} {2013})}\BibitemShut {NoStop}%
\bibitem [{\citenamefont {Bao}\ \emph {et~al.}(2011)\citenamefont {Bao},
  \citenamefont {Dong}, \citenamefont {Silverstein},\ and\ \citenamefont
  {Torroba}}]{nbao2011}%
  \BibitemOpen
  \bibfield  {author} {\bibinfo {author} {\bibfnamefont {N.}~\bibnamefont
  {Bao}}, \bibinfo {author} {\bibfnamefont {X.}~\bibnamefont {Dong}}, \bibinfo
  {author} {\bibfnamefont {E.}~\bibnamefont {Silverstein}}, \ and\ \bibinfo
  {author} {\bibfnamefont {G.}~\bibnamefont {Torroba}},\ }\href@noop {}
  {\bibfield  {journal} {\bibinfo  {journal} {J. High Energy Phys.}\ }\textbf
  {\bibinfo {volume} {10}},\ \bibinfo {pages} {123} (\bibinfo {year}
  {2011})}\BibitemShut {NoStop}%
\bibitem [{sup()}]{supinf}%
  \BibitemOpen
  \href@noop {} {\bibinfo  {journal} {See Supplemental Material at ...}\
  }\BibitemShut {NoStop}%
\bibitem [{\citenamefont {Baselmans}\ \emph {et~al.}(2012)\citenamefont
  {Baselmans}, \citenamefont {Yates}, \citenamefont {Diener},\ and\
  \citenamefont {de~Visser}}]{jbaselmans2012}%
  \BibitemOpen
\bibfield  {journal} {  }\bibfield  {author} {\bibinfo {author} {\bibfnamefont
  {J.}~\bibnamefont {Baselmans}}, \bibinfo {author} {\bibfnamefont
  {S.}~\bibnamefont {Yates}}, \bibinfo {author} {\bibfnamefont
  {P.}~\bibnamefont {Diener}}, \ and\ \bibinfo {author} {\bibfnamefont
  {P.}~\bibnamefont {de~Visser}},\ }\href@noop {} {\bibfield  {journal}
  {\bibinfo  {journal} {J. Low Temp. Phys.}\ }\textbf {\bibinfo {volume}
  {167}},\ \bibinfo {pages} {360} (\bibinfo {year} {2012})}\BibitemShut
  {NoStop}%
\bibitem [{\citenamefont {de~Visser}\ \emph {et~al.}(2010)\citenamefont
  {de~Visser}, \citenamefont {Withington},\ and\ \citenamefont
  {Goldie}}]{pdevisser2010}%
  \BibitemOpen
  \bibfield  {author} {\bibinfo {author} {\bibfnamefont {P.~J.}\ \bibnamefont
  {de~Visser}}, \bibinfo {author} {\bibfnamefont {S.}~\bibnamefont
  {Withington}}, \ and\ \bibinfo {author} {\bibfnamefont {D.~J.}\ \bibnamefont
  {Goldie}},\ }\href@noop {} {\bibfield  {journal} {\bibinfo  {journal} {J.
  Appl. Phys.}\ }\textbf {\bibinfo {volume} {108}},\ \bibinfo {pages} {114504}
  (\bibinfo {year} {2010})}\BibitemShut {NoStop}%
\bibitem [{\citenamefont {Swenson}\ \emph {et~al.}(2013)\citenamefont
  {Swenson}, \citenamefont {Day}, \citenamefont {Eom}, \citenamefont {Leduc},
  \citenamefont {Llombart}, \citenamefont {McKenney}, \citenamefont
  {Noroozian},\ and\ \citenamefont {Zmuidzinas}}]{lswenson2013}%
  \BibitemOpen
  \bibfield  {author} {\bibinfo {author} {\bibfnamefont {L.~J.}\ \bibnamefont
  {Swenson}}, \bibinfo {author} {\bibfnamefont {P.~K.}\ \bibnamefont {Day}},
  \bibinfo {author} {\bibfnamefont {B.~H.}\ \bibnamefont {Eom}}, \bibinfo
  {author} {\bibfnamefont {H.~G.}\ \bibnamefont {Leduc}}, \bibinfo {author}
  {\bibfnamefont {N.}~\bibnamefont {Llombart}}, \bibinfo {author}
  {\bibfnamefont {C.~M.}\ \bibnamefont {McKenney}}, \bibinfo {author}
  {\bibfnamefont {O.}~\bibnamefont {Noroozian}}, \ and\ \bibinfo {author}
  {\bibfnamefont {J.}~\bibnamefont {Zmuidzinas}},\ }\href {\doibase
  10.1063/1.4794808} {\bibfield  {journal} {\bibinfo  {journal} {J. Appl.
  Phys.}\ }\textbf {\bibinfo {volume} {113}},\ \bibinfo {eid} {104501}
  (\bibinfo {year} {2013})}\BibitemShut {NoStop}%
\bibitem [{Note2()}]{Note2}%
  \BibitemOpen
  \bibinfo {note} {At 64 mK, a $P_{read}$ of -64 (-100) dBm leads to a stored
  energy of 0.55 fJ (0.11 aJ), corresponding to 1.6$\times 10^8$ (3.1$\times
  10^4$) photons.}\BibitemShut {Stop}%
\bibitem [{\citenamefont {Nam}(1967)}]{snam1967}%
  \BibitemOpen
  \bibfield  {author} {\bibinfo {author} {\bibfnamefont {S.~B.}\ \bibnamefont
  {Nam}},\ }\href@noop {} {\bibfield  {journal} {\bibinfo  {journal} {Phys.
  Rev.}\ }\textbf {\bibinfo {volume} {156}},\ \bibinfo {pages} {470} (\bibinfo
  {year} {1967})}\BibitemShut {NoStop}%
\bibitem [{\citenamefont {Mooij}\ and\ \citenamefont
  {Klapwijk}(1983)}]{jmooij1983}%
  \BibitemOpen
  \bibfield  {author} {\bibinfo {author} {\bibfnamefont {J.~E.}\ \bibnamefont
  {Mooij}}\ and\ \bibinfo {author} {\bibfnamefont {T.~M.}\ \bibnamefont
  {Klapwijk}},\ }\href@noop {} {\bibfield  {journal} {\bibinfo  {journal}
  {Phys. Rev. B}\ }\textbf {\bibinfo {volume} {27}},\ \bibinfo {pages} {3054}
  (\bibinfo {year} {1983})}\BibitemShut {NoStop}%
\bibitem [{\citenamefont {Mattis}\ and\ \citenamefont
  {Bardeen}(1958)}]{dmattis1958}%
  \BibitemOpen
  \bibfield  {author} {\bibinfo {author} {\bibfnamefont {D.~C.}\ \bibnamefont
  {Mattis}}\ and\ \bibinfo {author} {\bibfnamefont {J.}~\bibnamefont
  {Bardeen}},\ }\href@noop {} {\bibfield  {journal} {\bibinfo  {journal} {Phys.
  Rev.}\ }\textbf {\bibinfo {volume} {111}},\ \bibinfo {pages} {412} (\bibinfo
  {year} {1958})}\BibitemShut {NoStop}%
\bibitem [{\citenamefont {Kaplan}(1979)}]{skaplan1979}%
  \BibitemOpen
  \bibfield  {author} {\bibinfo {author} {\bibfnamefont {S.~B.}\ \bibnamefont
  {Kaplan}},\ }\href@noop {} {\bibfield  {journal} {\bibinfo  {journal} {J. Low
  Temp. Phys.}\ }\textbf {\bibinfo {volume} {37}},\ \bibinfo {pages} {343}
  (\bibinfo {year} {1979})}\BibitemShut {NoStop}%
\bibitem [{\citenamefont {Gulian}\ and\ \citenamefont
  {Zharkov}(1982)}]{agulian1982}%
  \BibitemOpen
  \bibfield  {author} {\bibinfo {author} {\bibfnamefont {A.~M.}\ \bibnamefont
  {Gulian}}\ and\ \bibinfo {author} {\bibfnamefont {G.~F.}\ \bibnamefont
  {Zharkov}},\ }\href@noop {} {\bibfield  {journal} {\bibinfo  {journal} {J.
  Low Temp. Phys.}\ }\textbf {\bibinfo {volume} {48}},\ \bibinfo {pages} {125}
  (\bibinfo {year} {1982})}\BibitemShut {NoStop}%
\bibitem [{\citenamefont {Yassin}\ and\ \citenamefont
  {Withington}(1995)}]{gyassin1995}%
  \BibitemOpen
  \bibfield  {author} {\bibinfo {author} {\bibfnamefont {G.}~\bibnamefont
  {Yassin}}\ and\ \bibinfo {author} {\bibfnamefont {S.}~\bibnamefont
  {Withington}},\ }\href@noop {} {\bibfield  {journal} {\bibinfo  {journal} {J.
  Phys. D. Appl. Phys.}\ }\textbf {\bibinfo {volume} {28}},\ \bibinfo {pages}
  {1983} (\bibinfo {year} {1995})}\BibitemShut {NoStop}%
\bibitem [{\citenamefont {de~Visser}\ \emph {et~al.}(2011)\citenamefont
  {de~Visser}, \citenamefont {Baselmans}, \citenamefont {Diener}, \citenamefont
  {Yates}, \citenamefont {Endo},\ and\ \citenamefont
  {Klapwijk}}]{pdevisser2011}%
  \BibitemOpen
  \bibfield  {author} {\bibinfo {author} {\bibfnamefont {P.~J.}\ \bibnamefont
  {de~Visser}}, \bibinfo {author} {\bibfnamefont {J.~J.~A.}\ \bibnamefont
  {Baselmans}}, \bibinfo {author} {\bibfnamefont {P.}~\bibnamefont {Diener}},
  \bibinfo {author} {\bibfnamefont {S.~J.~C.}\ \bibnamefont {Yates}}, \bibinfo
  {author} {\bibfnamefont {A.}~\bibnamefont {Endo}}, \ and\ \bibinfo {author}
  {\bibfnamefont {T.~M.}\ \bibnamefont {Klapwijk}},\ }\href@noop {} {\bibfield
  {journal} {\bibinfo  {journal} {Phys. Rev. Lett.}\ }\textbf {\bibinfo
  {volume} {106}},\ \bibinfo {pages} {167004} (\bibinfo {year}
  {2011})}\BibitemShut {NoStop}%
\bibitem [{\citenamefont {Gao}\ \emph {et~al.}(2008)\citenamefont {Gao},
  \citenamefont {Zmuidzinas}, \citenamefont {Vayonakis}, \citenamefont {Day},
  \citenamefont {Mazin},\ and\ \citenamefont {Leduc}}]{jgao2008c}%
  \BibitemOpen
  \bibfield  {author} {\bibinfo {author} {\bibfnamefont {J.}~\bibnamefont
  {Gao}}, \bibinfo {author} {\bibfnamefont {J.}~\bibnamefont {Zmuidzinas}},
  \bibinfo {author} {\bibfnamefont {A.}~\bibnamefont {Vayonakis}}, \bibinfo
  {author} {\bibfnamefont {P.}~\bibnamefont {Day}}, \bibinfo {author}
  {\bibfnamefont {B.}~\bibnamefont {Mazin}}, \ and\ \bibinfo {author}
  {\bibfnamefont {H.}~\bibnamefont {Leduc}},\ }\href@noop {} {\bibfield
  {journal} {\bibinfo  {journal} {J. Low Temp. Phys.}\ }\textbf {\bibinfo
  {volume} {151}},\ \bibinfo {pages} {557} (\bibinfo {year}
  {2008})}\BibitemShut {NoStop}%
\bibitem [{\citenamefont {Martinis}\ \emph {et~al.}(2009)\citenamefont
  {Martinis}, \citenamefont {Ansmann},\ and\ \citenamefont
  {Aumentado}}]{jmartinis2009}%
  \BibitemOpen
  \bibfield  {author} {\bibinfo {author} {\bibfnamefont {J.~M.}\ \bibnamefont
  {Martinis}}, \bibinfo {author} {\bibfnamefont {M.}~\bibnamefont {Ansmann}}, \
  and\ \bibinfo {author} {\bibfnamefont {J.}~\bibnamefont {Aumentado}},\
  }\href@noop {} {\bibfield  {journal} {\bibinfo  {journal} {Phys. Rev. Lett.}\
  }\textbf {\bibinfo {volume} {103}},\ \bibinfo {pages} {097002} (\bibinfo
  {year} {2009})}\BibitemShut {NoStop}%
\bibitem [{\citenamefont {Catelani}\ \emph {et~al.}(2010)\citenamefont
  {Catelani}, \citenamefont {Glazman},\ and\ \citenamefont
  {Nagaev}}]{gcatelani2010}%
  \BibitemOpen
  \bibfield  {author} {\bibinfo {author} {\bibfnamefont {G.}~\bibnamefont
  {Catelani}}, \bibinfo {author} {\bibfnamefont {L.~I.}\ \bibnamefont
  {Glazman}}, \ and\ \bibinfo {author} {\bibfnamefont {K.~E.}\ \bibnamefont
  {Nagaev}},\ }\href@noop {} {\bibfield  {journal} {\bibinfo  {journal} {Phys.
  Rev. B}\ }\textbf {\bibinfo {volume} {82}},\ \bibinfo {pages} {134502}
  (\bibinfo {year} {2010})}\BibitemShut {NoStop}%
\bibitem [{\citenamefont {C\'{o}rcoles}\ \emph {et~al.}(2011)\citenamefont
  {C\'{o}rcoles}, \citenamefont {Chow}, \citenamefont {Gambetta}, \citenamefont
  {Rigetti}, \citenamefont {Rozen}, \citenamefont {Keefe}, \citenamefont
  {Rothwell}, \citenamefont {Ketchen},\ and\ \citenamefont
  {Steffen}}]{acorcoles2011}%
  \BibitemOpen
  \bibfield  {author} {\bibinfo {author} {\bibfnamefont {A.~D.}\ \bibnamefont
  {C\'{o}rcoles}}, \bibinfo {author} {\bibfnamefont {J.~M.}\ \bibnamefont
  {Chow}}, \bibinfo {author} {\bibfnamefont {J.~M.}\ \bibnamefont {Gambetta}},
  \bibinfo {author} {\bibfnamefont {C.}~\bibnamefont {Rigetti}}, \bibinfo
  {author} {\bibfnamefont {J.~R.}\ \bibnamefont {Rozen}}, \bibinfo {author}
  {\bibfnamefont {G.~A.}\ \bibnamefont {Keefe}}, \bibinfo {author}
  {\bibfnamefont {M.~B.}\ \bibnamefont {Rothwell}}, \bibinfo {author}
  {\bibfnamefont {M.~B.}\ \bibnamefont {Ketchen}}, \ and\ \bibinfo {author}
  {\bibfnamefont {M.}~\bibnamefont {Steffen}},\ }\href@noop {} {\bibfield
  {journal} {\bibinfo  {journal} {Appl. Phys. Lett.}\ }\textbf {\bibinfo
  {volume} {99}},\ \bibinfo {pages} {181906} (\bibinfo {year}
  {2011})}\BibitemShut {NoStop}%
\bibitem [{\citenamefont {Barends}\ \emph {et~al.}(2011)\citenamefont {Barends}
  \emph {et~al.}}]{rbarends2011}%
  \BibitemOpen
  \bibfield  {author} {\bibinfo {author} {\bibfnamefont {R.}~\bibnamefont
  {Barends}} \emph {et~al.},\ }\href@noop {} {\bibfield  {journal} {\bibinfo
  {journal} {Appl. Phys. Lett.}\ }\textbf {\bibinfo {volume} {99}},\ \bibinfo
  {pages} {113507} (\bibinfo {year} {2011})}\BibitemShut {NoStop}%
\bibitem [{\citenamefont {Lenander}\ \emph {et~al.}(2011)\citenamefont
  {Lenander} \emph {et~al.}}]{mlenander2011}%
  \BibitemOpen
  \bibfield  {author} {\bibinfo {author} {\bibfnamefont {M.}~\bibnamefont
  {Lenander}} \emph {et~al.},\ }\href@noop {} {\bibfield  {journal} {\bibinfo
  {journal} {Phys. Rev. B}\ }\textbf {\bibinfo {volume} {84}},\ \bibinfo
  {pages} {024501} (\bibinfo {year} {2011})}\BibitemShut {NoStop}%
\bibitem [{\citenamefont {Rainis}\ and\ \citenamefont
  {Loss}(2012)}]{drainis2012}%
  \BibitemOpen
  \bibfield  {author} {\bibinfo {author} {\bibfnamefont {D.}~\bibnamefont
  {Rainis}}\ and\ \bibinfo {author} {\bibfnamefont {D.}~\bibnamefont {Loss}},\
  }\href {\doibase 10.1103/PhysRevB.85.174533} {\bibfield  {journal} {\bibinfo
  {journal} {Phys. Rev. B}\ }\textbf {\bibinfo {volume} {85}},\ \bibinfo
  {pages} {174533} (\bibinfo {year} {2012})}\BibitemShut {NoStop}%
\bibitem [{\citenamefont {Wenner}\ \emph {et~al.}(2013)\citenamefont {Wenner}
  \emph {et~al.}}]{jwenner2013}%
  \BibitemOpen
  \bibfield  {author} {\bibinfo {author} {\bibfnamefont {J.}~\bibnamefont
  {Wenner}} \emph {et~al.},\ }\href {\doibase 10.1103/PhysRevLett.110.150502}
  {\bibfield  {journal} {\bibinfo  {journal} {Phys. Rev. Lett.}\ }\textbf
  {\bibinfo {volume} {110}},\ \bibinfo {pages} {150502} (\bibinfo {year}
  {2013})}\BibitemShut {NoStop}%
\bibitem [{\citenamefont {Kautz}(1978)}]{rkautz1978}%
  \BibitemOpen
  \bibfield  {author} {\bibinfo {author} {\bibfnamefont {R.~L.}\ \bibnamefont
  {Kautz}},\ }\href@noop {} {\bibfield  {journal} {\bibinfo  {journal} {J.
  Appl. Phys.}\ }\textbf {\bibinfo {volume} {49}},\ \bibinfo {pages} {308}
  (\bibinfo {year} {1978})}\BibitemShut {NoStop}%
\bibitem [{\citenamefont {Kaplan}\ \emph {et~al.}(1976)\citenamefont {Kaplan},
  \citenamefont {Chi}, \citenamefont {Langenberg}, \citenamefont {Chang},
  \citenamefont {Jafarey},\ and\ \citenamefont {Scalapino}}]{skaplan1976}%
  \BibitemOpen
  \bibfield  {author} {\bibinfo {author} {\bibfnamefont {S.~B.}\ \bibnamefont
  {Kaplan}}, \bibinfo {author} {\bibfnamefont {C.~C.}\ \bibnamefont {Chi}},
  \bibinfo {author} {\bibfnamefont {D.~N.}\ \bibnamefont {Langenberg}},
  \bibinfo {author} {\bibfnamefont {J.}~\bibnamefont {Chang}}, \bibinfo
  {author} {\bibfnamefont {S.}~\bibnamefont {Jafarey}}, \ and\ \bibinfo
  {author} {\bibfnamefont {D.~J.}\ \bibnamefont {Scalapino}},\ }\href@noop {}
  {\bibfield  {journal} {\bibinfo  {journal} {Phys. Rev. B}\ }\textbf {\bibinfo
  {volume} {14}},\ \bibinfo {pages} {4854} (\bibinfo {year}
  {1976})}\BibitemShut {NoStop}%
\end{thebibliography}

\begin{thebibliography}{11}%
\makeatletter
\providecommand \@ifxundefined [1]{%
 \@ifx{#1\undefined}
}%
\providecommand \@ifnum [1]{%
 \ifnum #1\expandafter \@firstoftwo
 \else \expandafter \@secondoftwo
 \fi
}%
\providecommand \@ifx [1]{%
 \ifx #1\expandafter \@firstoftwo
 \else \expandafter \@secondoftwo
 \fi
}%
\providecommand \natexlab [1]{#1}%
\providecommand \enquote  [1]{``#1''}%
\providecommand \bibnamefont  [1]{#1}%
\providecommand \bibfnamefont [1]{#1}%
\providecommand \citenamefont [1]{#1}%
\providecommand \href@noop [0]{\@secondoftwo}%
\providecommand \href [0]{\begingroup \@sanitize@url \@href}%
\providecommand \@href[1]{\@@startlink{#1}\@@href}%
\providecommand \@@href[1]{\endgroup#1\@@endlink}%
\providecommand \@sanitize@url [0]{\catcode `\\12\catcode `\$12\catcode
  `\&12\catcode `\#12\catcode `\^12\catcode `\_12\catcode `\%12\relax}%
\providecommand \@@startlink[1]{}%
\providecommand \@@endlink[0]{}%
\providecommand \url  [0]{\begingroup\@sanitize@url \@url }%
\providecommand \@url [1]{\endgroup\@href {#1}{\urlprefix }}%
\providecommand \urlprefix  [0]{URL }%
\providecommand \Eprint [0]{\href }%
\providecommand \doibase [0]{http://dx.doi.org/}%
\providecommand \selectlanguage [0]{\@gobble}%
\providecommand \bibinfo  [0]{\@secondoftwo}%
\providecommand \bibfield  [0]{\@secondoftwo}%
\providecommand \translation [1]{[#1]}%
\providecommand \BibitemOpen [0]{}%
\providecommand \bibitemStop [0]{}%
\providecommand \bibitemNoStop [0]{.\EOS\space}%
\providecommand \EOS [0]{\spacefactor3000\relax}%
\providecommand \BibitemShut  [1]{\csname bibitem#1\endcsname}%
\let\auto@bib@innerbib\@empty
\bibitem [{\citenamefont {de~Visser}\ \emph {et~al.}(2011)\citenamefont
  {de~Visser}, \citenamefont {Baselmans}, \citenamefont {Diener}, \citenamefont
  {Yates}, \citenamefont {Endo},\ and\ \citenamefont
  {Klapwijk}}]{pdevisser2011s}%
  \BibitemOpen
  \bibfield  {author} {\bibinfo {author} {\bibfnamefont {P.~J.}\ \bibnamefont
  {de~Visser}}, \bibinfo {author} {\bibfnamefont {J.~J.~A.}\ \bibnamefont
  {Baselmans}}, \bibinfo {author} {\bibfnamefont {P.}~\bibnamefont {Diener}},
  \bibinfo {author} {\bibfnamefont {S.~J.~C.}\ \bibnamefont {Yates}}, \bibinfo
  {author} {\bibfnamefont {A.}~\bibnamefont {Endo}}, \ and\ \bibinfo {author}
  {\bibfnamefont {T.~M.}\ \bibnamefont {Klapwijk}},\ }\href@noop {} {\bibfield
  {journal} {\bibinfo  {journal} {Phys. Rev. Lett.}\ }\textbf {\bibinfo
  {volume} {106}},\ \bibinfo {pages} {167004} (\bibinfo {year}
  {2011})}\BibitemShut {NoStop}%
\bibitem [{\citenamefont {Goldie}\ and\ \citenamefont
  {Withington}(2013)}]{dgoldie2013s}%
  \BibitemOpen
  \bibfield  {author} {\bibinfo {author} {\bibfnamefont {D.~J.}\ \bibnamefont
  {Goldie}}\ and\ \bibinfo {author} {\bibfnamefont {S.}~\bibnamefont
  {Withington}},\ }\href@noop {} {\bibfield  {journal} {\bibinfo  {journal}
  {Supercond. Sci. Technol.}\ }\textbf {\bibinfo {volume} {26}},\ \bibinfo
  {pages} {015004} (\bibinfo {year} {2013})}\BibitemShut {NoStop}%
\bibitem [{\citenamefont {Kaplan}(1979)}]{skaplan1979s}%
  \BibitemOpen
  \bibfield  {author} {\bibinfo {author} {\bibfnamefont {S.~B.}\ \bibnamefont
  {Kaplan}},\ }\href@noop {} {\bibfield  {journal} {\bibinfo  {journal} {J. Low
  Temp. Phys.}\ }\textbf {\bibinfo {volume} {37}},\ \bibinfo {pages} {343}
  (\bibinfo {year} {1979})}\BibitemShut {NoStop}%
\bibitem [{\citenamefont {Kaplan}\ \emph {et~al.}(1976)\citenamefont {Kaplan},
  \citenamefont {Chi}, \citenamefont {Langenberg}, \citenamefont {Chang},
  \citenamefont {Jafarey},\ and\ \citenamefont {Scalapino}}]{skaplan1976s}%
  \BibitemOpen
  \bibfield  {author} {\bibinfo {author} {\bibfnamefont {S.~B.}\ \bibnamefont
  {Kaplan}}, \bibinfo {author} {\bibfnamefont {C.~C.}\ \bibnamefont {Chi}},
  \bibinfo {author} {\bibfnamefont {D.~N.}\ \bibnamefont {Langenberg}},
  \bibinfo {author} {\bibfnamefont {J.}~\bibnamefont {Chang}}, \bibinfo
  {author} {\bibfnamefont {S.}~\bibnamefont {Jafarey}}, \ and\ \bibinfo
  {author} {\bibfnamefont {D.~J.}\ \bibnamefont {Scalapino}},\ }\href@noop {}
  {\bibfield  {journal} {\bibinfo  {journal} {Phys. Rev. B}\ }\textbf {\bibinfo
  {volume} {14}},\ \bibinfo {pages} {4854} (\bibinfo {year}
  {1976})}\BibitemShut {NoStop}%
\bibitem [{\citenamefont {Mazin}(2004)}]{bmazinphd}%
  \BibitemOpen
  \bibfield  {author} {\bibinfo {author} {\bibfnamefont {B.~A.}\ \bibnamefont
  {Mazin}},\ }\emph {\bibinfo {title} {Microwave kinetic inductance
  detectors}},\ \href@noop {} {Ph.D. thesis},\ \bibinfo  {school} {California
  Institute of Technology} (\bibinfo {year} {2004})\BibitemShut {NoStop}%
\bibitem [{\citenamefont {Barends}(2009)}]{rbarendsphd}%
  \BibitemOpen
  \bibfield  {author} {\bibinfo {author} {\bibfnamefont {R.}~\bibnamefont
  {Barends}},\ }\emph {\bibinfo {title} {Photon-detecting Superconducting
  Resonators}},\ \href@noop {} {Ph.D. thesis},\ \bibinfo  {school} {Delft
  University of Technology} (\bibinfo {year} {2009})\BibitemShut {NoStop}%
\bibitem [{\citenamefont {Zmuidzinas}(2012)}]{jzmuidzinas2012s}%
  \BibitemOpen
  \bibfield  {author} {\bibinfo {author} {\bibfnamefont {J.}~\bibnamefont
  {Zmuidzinas}},\ }\href@noop {} {\bibfield  {journal} {\bibinfo  {journal}
  {Ann. Rev. Condens. Matter Phys.}\ }\textbf {\bibinfo {volume} {3}},\
  \bibinfo {pages} {169} (\bibinfo {year} {2012})}\BibitemShut {NoStop}%
\bibitem [{\citenamefont {Pozar}(1998)}]{pozar}%
  \BibitemOpen
  \bibfield  {author} {\bibinfo {author} {\bibfnamefont {D.~M.}\ \bibnamefont
  {Pozar}},\ }\href@noop {} {\emph {\bibinfo {title} {Microwave
  {E}ngineering}}},\ \bibinfo {edition} {2nd}\ ed.\ (\bibinfo  {publisher}
  {John Wiley \& Sons, Inc.},\ \bibinfo {year} {1998})\BibitemShut {NoStop}%
\bibitem [{\citenamefont {Yassin}\ and\ \citenamefont
  {Withington}(1995)}]{gyassin1995s}%
  \BibitemOpen
  \bibfield  {author} {\bibinfo {author} {\bibfnamefont {G.}~\bibnamefont
  {Yassin}}\ and\ \bibinfo {author} {\bibfnamefont {S.}~\bibnamefont
  {Withington}},\ }\href@noop {} {\bibfield  {journal} {\bibinfo  {journal} {J.
  Phys. D. Appl. Phys.}\ }\textbf {\bibinfo {volume} {28}},\ \bibinfo {pages}
  {1983} (\bibinfo {year} {1995})}\BibitemShut {NoStop}%
\bibitem [{\citenamefont {Clem}(2013)}]{jclem2013}%
  \BibitemOpen
  \bibfield  {author} {\bibinfo {author} {\bibfnamefont {J.~R.}\ \bibnamefont
  {Clem}},\ }\href@noop {} {\bibfield  {journal} {\bibinfo  {journal} {J. Appl.
  Phys}\ }\textbf {\bibinfo {volume} {113}},\ \bibinfo {pages} {013910}
  (\bibinfo {year} {2013})}\BibitemShut {NoStop}%
\bibitem [{\citenamefont {Holloway}\ and\ \citenamefont
  {Kuester}(1995)}]{cholloway1995}%
  \BibitemOpen
  \bibfield  {author} {\bibinfo {author} {\bibfnamefont {C.~L.}\ \bibnamefont
  {Holloway}}\ and\ \bibinfo {author} {\bibfnamefont {E.~F.}\ \bibnamefont
  {Kuester}},\ }\href@noop {} {\bibfield  {journal} {\bibinfo  {journal} {IEEE
  Trans. on Micr. Theory and Tech.}\ }\textbf {\bibinfo {volume} {43}},\
  \bibinfo {pages} {2695} (\bibinfo {year} {1995})}\BibitemShut {NoStop}%
\end{thebibliography}
\end{document}